\documentclass[a4paper,11pt]{article}
\usepackage{jinstpub} % for details on the use of the package, please see the JINST-author-manual
\notoc
\usepackage{lineno}
\usepackage[acronym]{glossaries}
\glsdisablehyper

\newacronym{idaq}{IDAQ}{INFN Data Acquisition}
\newacronym{sc}{SC}{Slow-control}
\newacronym{ci}{CI}{Continuous Integration}
\newacronym{ber}{BER}{Bit Error Rate}
\newacronym[plural=DAQs,firstplural=Data Acquisition systems]{daq}{DAQ}{Data Acquisition system}
\newacronym[plural=ASICs,firstplural=Application-Specific Integrated Circuits]{asic}{ASIC}{Application-Specific Integrated Circuit}
\newacronym[plural=FPGAs,firstplural=Field-Programmable Gate Arrays]{fpga}{FPGA}{Field Programmable Gate Array}

\title{A flexible electronics and DAQ system for the Timepix4 ASIC}

\author[a,1]{N.~V.~Biesuz\note{Corresponding author.}}
\author[c]{J.~Alozy}
\author[c]{R.~Bolzonella}
\author[a,b]{V.~Cavallini}
\author[a]{A.~Cotta~Ramusino}
\author[a,b]{M.~Fiorini}
\author[c]{X.~Llopart Cudie}
\author[a,b]{A.~Saputi}

 \affiliation[a]{INFN sezione di Ferrara,\\via Saragat 1 - 44122 Ferrara, Italy}
 \affiliation[b]{Universit\'a di Ferrara,\\via Ludovico Ariosto, 35 - 44121 Ferrara, ITALY}
 \affiliation[c]{CERN,\\Esplanade des Particules 1 - P.O. Box 1211 Geneva 23 Switzerland, SWITZERLAND}

\emailAdd{biesuz@fe.infn.it}

\abstract{
The INFN Data Acquisition system (IDAQ) is designed for data-driven readout of the Timepix4 ASIC, supporting simultaneous time-of-arrival and time-over-threshold measurements with nanosecond precision. The system architecture combines dedicated hardware for high-speed serial data transmission, FPGA-based data processing, and a Trigger Logic Unit for synchronization and timing distribution. It operates in continuous streaming and triggered modes, ensuring low-latency and high-throughput performance. The system has been deployed in test-beam campaigns within the 4DPHOTON project, demonstrating stable operation, precise timing alignment, and reliable data handling under high-rate conditions.
}


\begin{document}
\maketitle
\flushbottom
%===========================================================
\section{Introduction}\label{sec:introduction}
%===========================================================

\begin{figure}[htb]
\centering
\includegraphics[width=0.5\textwidth]{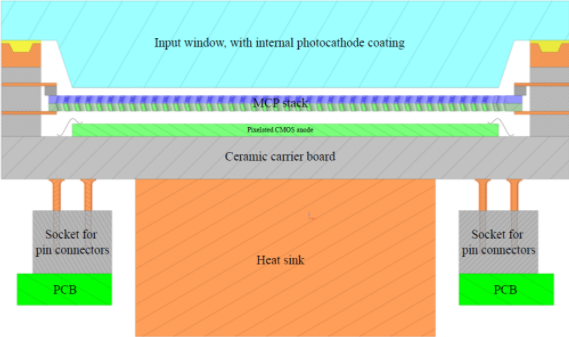}
\caption{Schematic cross section view of the 4DPHOTON detector, taken from~\cite{Alozy2022SinglePhoton}.}
\label{fig:tube_cross_section}
\end{figure}

The 4DPHOTON project develops a single-photon imaging detector combining high spatial granularity with picosecond-level timing. Its hybrid architecture couples a photocathode, an MCP amplification stack, and a Timepix4 readout anode~\cite{Fiorini2018,Alozy2022SinglePhoton}, as shown in Figure~\ref{fig:tube_cross_section}. The ASIC capabilities are described in Section~\ref{sec:tpx4_mpx4_asics}.

A dedicated, modular \gls{daq} interfaces Timepix4, with provision for extension to Medipix4. Standard high-speed links and Ethernet transport support cost-effective deployment in laboratory and test-beam environments.

The same \gls{daq} also supports other INFN projects using solid-state sensors for medical and cultural heritage applications, beam diagnostics, and fluorescence microscopy.

\section{The Timepix4 ASIC}\label{sec:tpx4_mpx4_asics}

Timepix4 is a hybrid pixel readout \gls{asic} developed by the CERN Medipix Collaboration in 65~nm CMOS technology. Its $448\times512$ pixel matrix has a $55~\mathrm{\mu m}$ pitch. Each pixel integrates a charge-sensitive front-end, discriminator, and Time-to-Digital Converter for simultaneous Time-of-Arrival and Time-over-Threshold measurements~\cite{Llopart2022Timepix4,Ballabriga2023Timepix4AFE}.

Data-driven readout produces 64-bit event words for hit pixels, with 195~ps timestamp binning and support for hit rates of several $\mathrm{MHz/mm^2}$. Up to 16 serial links operate between 40~Mb/s and 10.24~Gb/s, using 64b/66b encoding. The number of active links and their rates can be tailored to the required throughput, simplifying off-chip electronics.

Acquisition is controlled by \texttt{T0\_SYNC}, which resets the global time reference, and \texttt{SHUTTER}, which enables data taking. Configuration and monitoring use a dedicated \gls{sc} interface.
The \gls{sc} interface provides separate input and output buses, each comprising LVDS serial-data and clock signals.

Medipix4~\cite{Sriskaran2024Medipix4} shares the same \gls{sc} interface and uses exclusively frame-based readout. It features a $320\times320$ pixel matrix with a $75~\mathrm{\mu m}$ square pitch and two thresholds per pixel. Data are transmitted over four synchronous DDR links operating at 20--640~Mb/s per link.

\section{Timepix4 Data Acquisition system: state of the Art}

Existing Timepix4 \gls{daq} platforms offer different throughput and hardware requirements. 
SPIDR4~\cite{Heijden2017} supports acquisitions at the maximum data rate of 163.84~Gb/s but requires dedicated acquisition hardware to exploit its full bandwidth; in its reduced-bandwidth configuration it provides a maximum data rate of 5.12~Gb/s that can constrain measurements.
Katherine~\cite{KatherineTPX4} provides an alternative for laboratory and test-beam use. This set-up provides a maximum reported throughput of approximately $350~\mathrm{Mhits/s}$ equivalent to 23.1~Gb/s using its PCIe interface. While this is comparable to the thoughput of the described system this feature is still in development and became available too late for this project.

This motivates the development of a scalable \gls{daq} framework for Timepix4, called \gls{idaq} specifically targeting detector development and test-beam campaigns, with a target aggregate throughput of 10-20~Gb/s and provision for future extension to Medipix4.
The main advantage of \gls{idaq} is low-cost deployment for laboratory measurements and test-beam campaigns using commercial FPGA hardware and standard Ethernet interfaces, with provision for extension to Medipix4.

%===========================================================
\section{IDAQ Hardware system description}\label{sec:idaq_arch}
%===========================================================

The \gls{idaq} architecture is based on a modular hardware stack composed of:
\begin{itemize}
  \item an \gls{idaq} control board;
  \item an \gls{idaq} FMC adapter card;
  \item an \gls{asic}-specific chipboard hosting the detector;
\end{itemize}

The \gls{idaq} system connects the front-end electronics to remote acquisition servers. Its architecture addresses the requirements of moderate throughput, low hardware cost, and ease of development by using the commercial KCU105 FPGA evaluation board. This choice limits the input bandwidth to eight links operating at up to 10.24~Gb/s each, corresponding to a peak aggregate input rate of 81.92~Gb/s. Mechanical constraints of the MCP-PMT design restrict signal routing, limiting reliable operation to 2.56~Gb/s per link at a \gls{ber} below $10^{-12}$. The aggregate input bandwidth is 20.48~Gb/s, corresponding to a maximum payload rate of 19.85~Gb/s after 64b/66b encoding overhead. Data transmission via UDP over 10~Gb Ethernet enables acquisition through standard network interface cards, limiting server-side costs. The two 10~GbE outputs available theoretically sustain 19.14~Gb/s with a standard MTU or 19.85~Gb/s with a 9000-byte MTU, before implementation overhead. The measured aggregate throughput is 17.7~Gb/s; sustained inputs above this rate therefore require data reduction, while buffering accommodates short bursts. 

The \gls{asic}-specific chipboard hosts the detector and provides electrical interconnections and low-noise power through on-board regulators, following the SPIDR4 chipboard concept for solid-state sensors. Compatibility with existing chipboards facilitates the development of test-beam trackers and extends the range of applications to solid-state sensor characterization and tomography. A dedicated chipboard was also developed for the 4DPHOTON project to support MCP-PMT-based detectors.

A simple, low-cost FMC adapter card connects the KCU105 to the chipboard, providing level shifting, signal buffering, and clock-handling circuitry. This modular interface supports compatibility with existing detector hardware, including the chipboard developed for the SPIDR4 system.

%===========================================================
\section{Gateware for Timepix4}\label{sec:fw_tpx4}
%===========================================================

\subsection{Gateware distribution and versioning}\label{sec:hog}

The gateware is hosted in a GitLab repository containing source files and AMD Vivado project-generation recipes. Hog~\cite{Hog2023} manages project generation and compilation, both locally and through GitLab \gls{ci}~\cite{GitLabCI}, ensuring reproducibility and traceability to the exact source version.

Development uses a stable main branch and feature branches. Merge requests trigger a \gls{ci} pipeline that generates a clean project, runs synthesis, implementation and self-checking tests, and builds documentation. Successful merges automatically produce tagged releases containing the compiled gateware, version information, compilation reports, helper scripts, ChangeLog, documentation and XML address maps.

Releases are distributed to the collaboration under the CERN Open Hardware Licence. An automated REUSE~\cite{REUSEsoftware} check verifies copyright and licence information in all source files.

\subsection{Gateware architecture}\label{sec:fw_arch}

\begin{figure}
    \centering
    \includegraphics[width=0.5\linewidth]{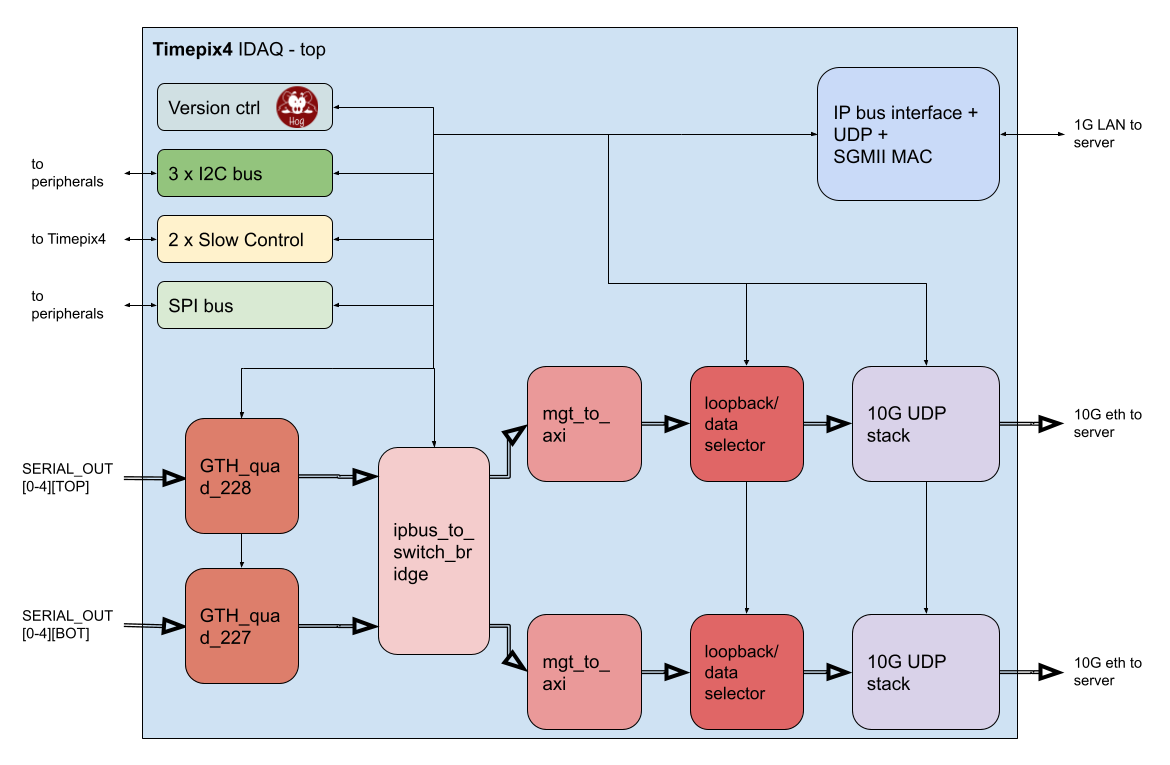}
    \caption{\gls{idaq} gateware architecture. Identical entities share the same colour; single and double arrows indicate control and high-speed data paths, respectively.}
    \label{fig:cbf_top}
\end{figure}

The XCKU040 gateware separates control and data processing, as shown in Figure~\ref{fig:cbf_top}. The control path configures and monitors the system through IPbus over 1~GbE, while two independent data paths connect the Timepix4 half-matrices to the two 10~GbE outputs. Configuration changes to the data paths are disabled during acquisition.

The control interface uses a standard UDP-IP stack to forward frames to an IPbus control module, which converts them into IPbus transactions~\cite{IPbus2015}. Registers and memories are organised hierarchically, with each functional module connected through an IPbus bridge. These modules provide version and status monitoring, board configuration, I\textsuperscript{2}C and SPI access, \gls{asic} slow control, and data-path configuration. XML address maps distributed with each release support the control software and automatically generate VHDL address-decoding packages. Expected version values are embedded during GitLab \gls{ci} project generation to support gateware--software consistency checks.

Two GTH quads receive up to eight Timepix4 links. A dedicated wrapper handles transceiver initialisation, block-lock detection, 64b/66b descrambling, and clock-domain crossing to the UDP processing clock. Each link includes two 512-word, 64-bit FIFOs; additional FIFOs buffer subsequent processing stages. Predefined transceiver configurations support line rates of 2.56~Gb/s. A gateware with configurable link speed of 1.24, 2.56 and 5.12~Gb/s is under development.

The data paths do not propagate back-pressure to the \gls{asic}. Internal FIFOs absorb short bursts above the sustained output bandwidth; prolonged overload causes data loss, monitored through dedicated registers. Each output uses the open-source \textit{verilog-ethernet} UDP/IP stack~\cite{VerilogEthernet}, supporting 1500-byte MTU packets. The \texttt{mgt\_to\_axi} module packetises the incoming stream and converts it to AXI-Stream at 156.25~MHz, introducing three clock cycles between consecutive blocks. Partial packets are also transmitted to avoid prolonged data retention.

For each output, \texttt{ipbus\_to\_switch\_bridge} either selects one link or merges four links operating at 2.56~Gb/s each. Single-link selection is fixed during acquisition and introduces no buffering or back-pressure. The merging implementation buffers incoming data, removes replicated control words, and monitors words dropped during congestion. Enabling all four links accommodates bursty inputs, while masking links reduces the aggregate rate for sustained acquisition. This merging stage also provides an extension point for application-specific data reduction. Hardware validation of the data path is presented in Section~\ref{sec:performance}.

\subsection{Application-specific data reduction}

The \texttt{ipbus\_to\_switch\_bridge} receives all input data streams, providing a natural integration point for project-specific data reduction, such as clustering, masking, or aggregation. A module template provides input/output buffering and IPbus integration, exposing a dedicated bus for each IPbus node. Processing runs synchronously with the UDP clock at 156.25~MHz.

Users can adopt the default 4:1 data merger or implement custom processing within the template. The existing \gls{ci} workflow automates gateware builds and validation, simplifying the selection and deployment of project-specific variants. A clustering implementation is currently under development as an additional gateware option.

\section{Extension to Medipix4}\label{sec:fw_mpx4}

Support for the frame-based Medipix4 readout reuses the control path and UDP outputs, replacing the Timepix4 receiver and merging stages with sixteen DDR interfaces operating at 20--320~MHz. Dedicated deserializers detect frames, decode headers, and reconstruct 64-bit data streams for transmission over two 10~GbE links. Migration to the lower-cost AXKU040 platform is ongoing, targeting simultaneous readout of two Medipix4 ASICs through dual FMC-LPC interfaces and four 10~GbE links.

\section{Performance evaluation}\label{sec:performance}

The sustained data-path throughput was measured using a dedicated gateware test harness feeding counter-generated data through \texttt{mgt\_to\_axi} and the 10~GbE UDP stack. Over approximately 13~s, exceeding a typical test-beam spill duration, each output sustained $8.85\pm0.05~\mathrm{Gb/s}$, yielding an aggregate throughput of $17.7~\mathrm{Gb/s}$. The difference from the ideal UDP payload throughput is attributed to packetization and implementation overhead.

Link reliability was evaluated using PRBS-31 sequences transmitted simultaneously on all Timepix4 links and checked by the FPGA receivers. With the 4DPHOTON Tube card, no errors were observed during 10~s tests at 2.56~Gb/s per link, corresponding to a tested \gls{ber} sensitivity of $3.9\times10^{-11}$ per link. Repeated reads of a 256-bit Timepix4 register yielded a \gls{sc} transaction rate of $13\pm0.5~\mathrm{kOps/s}$.

%===========================================================
\section{Conclusions}\label{sec:conclusion}
%===========================================================

A flexible and modular \gls{daq} system for the Timepix4\ \gls{asic} has been developed within the 4DPHOTON project. 
The system comprises the complete acquisition chain designed to support full characterization of the 4DPHOTON detector and dedicated beam-test campaigns.

The achieved performance enables complete detector testing and has been validated through device acceptance tests and beam-test measurements. 
System throughput and timing performance are primarily determined by the hardware architecture; further improvements would require custom hardware or application-specific optimizations.

\section{Outlook and Future Developments}

Future work will improve portability and maintainability by replacing vendor-specific IP with open-source components from Open Logic~\cite{OpenLogic} and \textit{verilog-ethernet}.
Further gateware improvements will include configurable input-link speeds and the integration of application-specific data-reduction algorithms.

Support for Medipix4 is under development, reusing the control path with a dedicated data path. A port to the Alinx AXKU040 board is also planned: its four 10~GbE interfaces will increase output bandwidth, while retaining the XCKU040 FPGA and core firmware architecture.

\acknowledgments

This work was carried out in the context of the Medipix4 Collaboration based at CERN, and in the framework of the Medipix4 project funded by INFN CSN5. This project has received funding from the European Research Council (ERC) under the European Union’s Horizon 2020 research and innovation programme (Grant agreement No. 819627, 4DPHOTON project).

\bibliographystyle{iopart-num}

\end{document}